# Can Background Baroque Music Help to Improve theMemorability of Graphical Passwords?

Haichang Gao[1,2], Xiuling Chang[1], Zhongjie Ren[1],
Uwe Aickelin[2], and Liming Wang[1]

1 Software Engineering Institute, Xidian University, Xi'an, Shaanxi 710071, P.R. China
2 School of Computer Science, The University of Nottingham, Nottingham, NG8 1BB, U.K. `hchgao@xidian.edu.cn`

**Abstract.** Graphical passwords have been proposed as an alternative to alphanumeric passwords with their advantages in usability and security. However, they still tend to follow predictable patterns that are easier for attackers to exploit, probably due to users' memory limitations. Various literatures show that baroque music has positive effects on human learning and memorizing. To alleviate users' memory burden, we investigate the novel idea of introducing baroque music to graphical password schemes (specifically DAS, PassPoints and Story) and conduct a laboratory study to see whether it is helpful. In a ten minutes short-term recall, we found that participants in all conditions had high recall success rates that were not statistically different from each other. After one week, the music group coped PassPoints passwords significantly better than the group without music. But there was no statistical difference between two groups in recalling DAS passwords or Story passwords. Furthermore, we found that the music group tended to set significantly more complicated PassPoints passwords but less complicated DAS passwords.

## 1 Introduction

Graphical passwords have been proposed as an alternative to alphanumeric passwords and the main motivation is the hypothesis that people perform far better when remembering pictures rather than words [1, 2]. Visual objects seem to offer a much larger set of usable passwords. It is conceivable that humans would be able to remember stronger passwords of a graphical nature. However, users still tend to choose passwords that are memorable in some way, which means that the graphical passwords still tend to follow predictable patterns that are easier for attackers to exploit [6, 13, 14].

Various literatures reveal that users are the 'weakest link' in any password authentication mechanism, probably due to their memory limitations [11]. Although human memory capacity is unlikely to increase significantly over the next few years, recent psychological and physiological studies indicate that certain music like baroque music has positive effects of great importance on human memorizing and learning [20, 22].

Motivated by these observations, we investigate the novel idea of introducing background baroque music to graphical password schemes with the purpose of alleviating users' memory burden and improving usable security. Based on DAS, PassPoints and Story schemes, we conduct a laboratory study to explore the efficiency of background baroque music on memorizing graphical passwords. We are also interested in whether the background music would enable users to choose more complicated or less predictable passwords, which are usually more resistant to dictionary and other guess attacks.

The following section briefly reviews graphical password schemes and related works. Sections 3 and 4 describe the methodology of our studies and present the results respectively. Section 5 provides several interpretations to the experiment and discusses the experimental results. Conclusion and future work are addressed in Section 6.

## 2 Related Works

### 2.1 Graphical Password Schemes

In the open literature to date, the ubiquity of graphical interfaces for applications and input devices, such as the mouse, stylus and touch-screen, has enabled the emergence of graphical authentications. There have been three dominant techniques available which can be defined as: Drawmetrics (DAS [4], Syukri [9], YAGP [21]), Locimetrics (Blonder [3], PassPoints [7]) and Cognometrics (Déjà Vu [5], Story [6], Passfaces [10]) [19].

Drawmetrics systems require users to reproduce a pre-drawn outline drawing on a grid. A well-known scheme in this category is DAS which liberates users from remembering complicate text strings and has the advantage of better security over alphanumerical passwords [4]. Nevertheless, Passdoodle revealed that people are able to remember complete doodle images while less likely to recall the stroke order [8]. Furthermore, Thorpe and Van Oorschot found that users tend to design symmetrical and centered or approximately centered passwords, significantly reducing password space in practice and impacting the security [14]. Gao et al. proposed a modification to DAS where approximately correct drawings can be accepted, based on Levenshtein distance string matching and "trend quadrants" looking at the direction of strokes [21].

Locimetrics systems are based on the method of loci, an old and well-known mnemonic [18]. Originating in Blonder's work, the approach involves users choosing several sequential locations in an image [3]. PassPoints [7] is a representative scheme of this category, where users may choose any place in the image as a password click point. Since it is a cue of great importance for users to recall their passwords, the image should be complex and visually rich enough to have many potentially memorable click points. This scheme was found that although relatively usable, security concerns remains. A primary security problem is hotspots: people tend to select obvious points in the image with high visual salience, leading to a reduced effective password space that facilitates more successful dictionary attacks [12,13].

In the Cognometrics systems, users must recognize the target images embedded amongst a set of distractor images. This category includes Passfaces which relies on face recognition [10], Déjà Vu [5] based on abstract images and Story [6] where users are suggested to create a story and so on. User studies by Valentine have shown that Passfaces has a high degree of memorability [15, 16], but Davis found that people tended to select faces of their own race and gender [6]. Assigning faces to users arbitrarily may alleviate the problem, whereas it would lead people hard to remember the password. A similar scheme to Passfaces is Story where the password selection is sufficiently free from bias [6]. But, the Story is not as good as Passfaces in memorability, because few people actually choose stories despite the suggestion. In addition, memorability for abstract images in Déjà Vu was found to be only half as good as that for photographic images with a clear central subject [17].

Through the above discussion, we find that most graphical passwords either tend to follow predictable patterns or have a low degree of memorability. The crux of the problem is the users' memory limitations. As human memory capacity is unlikely to increase significantly over the next few years, creating a nice environment for memorizing passwords might alleviate users' burden. There are demonstrations that music can improve memory and in what flows we will illustrate it.

### 2.2 The Efficiency of Baroque Music

Extensive researches have shown that music has different uses for education and therapy [20]. As our particular interest is to explore the role of music in learning and memorizing graphical passwords, we will briefly review the researches into the effects of music on learning in this subsection.

Georgi Lozanov, a Bulgarian psychologist, made remarkable impact in integrating music into teaching practice. He created a teaching method called 'Suggestopedia', wherein the use of background music, particularly the baroque music with a rate of 50 to 70 beats per minute (BPM), is a cornerstone of accelerated learning techniques. It is stated that the method of Suggestopedia involves three stages where different types of music are used for specific purposes. First, introduce music to relax participants and help them to achieve the optimum state for learning. Second, listen to an "active concert" with music from Mozart, Beethoven and Brahms. Finally, apply a "passive concert" to help participants move the information into the long-term memory. While no details are given as to which exact music is suggested for the first stage, both the concerts in the latter two stages result in high memory retention [22]. Furthermore, Lozanov says that "well organized Suggestopedia accelerates learning 5 times on an average" [22].

Baroque music can help the brain produce alpha waves, and information imbued with music has a greater likelihood of being encoded in the long-term memory by the brain. That is why accelerated learning techniques introduce music into the learning process. For example, 'Mozart Effect' [23] is a phenomenon that music has a positive effect on learning and memory. In the following sections, we bring background baroque music to graphical password schemes, specifically, PassPoints, DAS and Story, and do an investigation to check whether it can improve users' memory or induce users to set stronger passwords.

## 3 User Study

As mentioned earlier, our evaluation is based on three representative graphical password schemes. For the purpose of collecting and analyzing the success rate, user habits, and login time automatically, we reproduce three schemes which are intentionally very closely modeled after DAS [4], PassPoints [7] and Story [6], respectively. We still adopt the names "DAS", "PassPoints" and "Story" for convenience. In this section, after describing the three schemes deployed in our experiments, we will present our methodology in great detail.

## 3.1 Brief Introduction of the Reproduced Schemes

DAS is a drawing reproduction based scheme, where a $5 \times 5$ grid was deployed for users to draw on. Each grid cell is denoted by rectangular discrete coordinates (x, y) [0, 4] × [0, 4]. A completed drawing is encoded as the ordered sequence of cells that the user crosses whilst constructing the secret, with a distinguished coordinate pair (5, 5) inserted in both ends of each stroke. Two passwords are identical if the encoding is the same. Figure 1 shows how DAS works. Input a graphical password consisting of three strokes, which are colored by black, green and red in sequence. The drawing is mapped to (5,5)(1,2)(1,3)(2,3)(3,3)(3,2)(5,5); (5,5)(2,1)(2,2)(2,3)(5,5); (5,5)(2,1)(5,5).

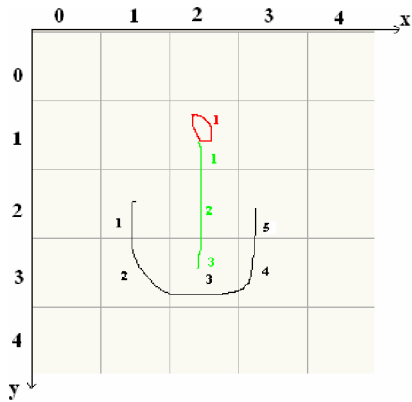

**Fig. 1.** An example of DAS password with length being 9

In the PassPoints scheme, users are required to select several positions in a single image as their passwords and click close to the chosen points in correct order and within a tolerance distance for authentication. For example, the password in Figure 2 contains five click points orderly labeled by small red rectangle.

In Story, a password is a sequence of k (k9) images selected by the user to make a "story". To keep consistent with that in [6], the images used here are also classified into nine categories, which are animals, cars, women, food, children, men, objects, nature, and sports. Images of "men" and "women" are gathered from [FordModels.com](FordModels.com) and the others [http://images.google.com](http://images.google.com). Figure 3 shows the interface of Story, where the man, woman, car and the house are orderly selected and the underlying story is "a gentle man and his girlfriend drive a car to their house".

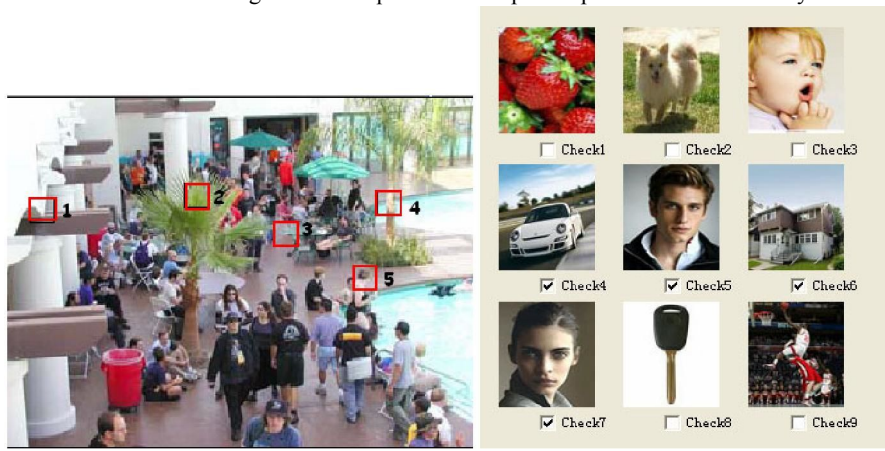

**Fig. 2.** Passwords in PassPoints with length being 5   **Fig. 3.** An example of Story password

## 3.2 Experiments

We conducted a lab study with 28 subjects (16 males and 12 females). All the subjects were university students of computer science and in the age range of 20 to 30. We hypothesized that background music could improve humans memory and then induced people to choose more complex passwords and take less time to log in. This study used a between-subjects design and had two conditions; half of the subjects were assigned to the control group (without background music) and half to the music group. None of them had previously used DAS, PassPoints or Story passwords. We chose the baroque music suggested by Lozanov with a rate of 50 to 70 BMP as the background music and utilized a Lenovo speaker to play it. The volume was set to 3040 decibels as suggested.

Our study included two lab-based sessions. Session 1 took about two hours. At the beginning of Session 1, each participant was asked to read an instruction document. This provided information of their activities on the experiments and helped them know how DAS, PassPoints and Story work. To make the rules clearer, an example was included in each scheme. Then participants were required to complete the registration and login of DAS, then PassPoints, and finally Story. People were asked to reenter the password to confirm it. After a short delay (about 10 minutes), participants were asked to log in within three attempts. In the end, participants need answer a demographic questionnaire collecting information including age, sex and experience on graphical passwords.

One week later, at Session 2, all the participants returned to the lab and tried to log in each scheme within three attempts using their previously created passwords.

## 4 Results

We used two types of statistical tests to assess whether differences in the data reflect actual differences between conditions or whether these may have occurred by chance. A t-test (two tails) was used for comparing the means of two groups and Fisher's xact test was used to compare recall success rates. In all cases, we regard a value of P<0.20 as indicating that the groups being tested are different from each other with at least

80% probability, making the result statistically significant. In the tables, "not significant" indicates that the test revealed no statistically significant difference between the two conditions (i.e., P>0.20).

### 4.1 Success Rates

We first examine success rates as a measure of participants' performance. Table 1 compares the successful recalls in each group.

Table 1. Success rates in each group for DAS, PassPoints and Story

| Group | 10-minute test | | 1-week test | |
|---|---|---|---|---|
| | ratio | Fisher-test | ratio | Fisher-test |
| DAS (no music) | 78.6% | P=0.59 | 71.4% | P=1 |
| DAS (music) | 92.9% | | 64.3% | |
| PassPoints (no music) | 100% | P=1 | 35.7% | P=.004 |
| PassPoints (music) | 100% | | 92.9% | |
| Story (no music) | 100% | P=1 | 92.9% | P=1 |
| Story (music) | 100% | | 92.9% | |

In the 10 minutes short term phase, the success rates were high on the whole, indicating that participants' memory was not strongly taxed. In PassPoints and Story, participants under both conditions recalled their passwords. In DAS, the success rate of the music group was 92.9%, higher than that of the control one (78.6%). However, a Fisher's exact test yields a result of P=0.59, indicating that the difference was not statistically significant.

Table 2. Complexity of DAS secrets

| Group | | DAS (no music) | DAS (music) |
|---|---|---|---|
| Strokes | Avg. t-test | 3.36 | 3.71 |
| | | Not significant | |
| | S.d. | 1.71 | 2.25 |
| | Max | 7 | 7 |
| | Min | 1 | 1 |
| Password Length | Avg. t-test | 13.79 | 10.43 |
| | | t=1.34, P<0.20 | |
| | S.d. | 6.39 | 6.41 |
| | Max | 27 | 21 |
| | Min | 2 | 1 |

After one week, the performances of two groups varied in schemes. Both groups in Story had the same success recall rate 92.9%, but differed in DAS and PassPoints. In DAS, only 64.3% of the music group and 71.4% of the control group were able to recall their passwords. It appears that the control group performed better than the music group. It should be noted that it was only a difference of one person in practice. The result of Fisher's exact test showed that there was no statistical difference between two conditions. In PassPoints, we found a significant difference between two groups. The

Can Background Baroque Music Help to Improve the Memorability?music group was significantly more likely to successfully recall the passwords than the control group. In addition, the success rate of the control group decreased from 100% in the previous phase to 35.7% while the success rate of the music group only decreased by 7.1%. It aligns with psychology research which continues to show that certain music advance the long-term memory.

The results suggest that the background music works differently when it was available in different graphical password schemes. In Drawmetrics and Cognometrics systems, background music seems to have no influence on short-recalls or long-term memory. But in Locimetrics systems, it appears that background music could significantly help people remember passwords in long-term memory.

### 4.2 Password Complexity

For each scheme, we compare password complexity in both groups. While the password length in PassPoints or Story is easy to understand, it is necessary to explain it in DAS. In DAS, the length of a password yields by adding the lengths of its component strokes wherein the length of a stroke is the number of coordinate pairs it contains exclusive of the distinguished ones(5,5). For example, for the password in Figure2, the length of each stroke is 5, 3 and 1 respectively, producing a password length of 9.

**Table 3.** Complexity of PassPoints and Story secrets

| Group | Password length | | | | |
|---|---|---|---|---|---|
| | Avg. | t-test | S.d. | Max | Min |
| PassPoints (no music) | 3.79 | t=1.61, P<0.20 | 1.20 | 5 | 1 |
| PassPoints (music) | 4.5 | | 1.05 | 6 | 3 |
| Story (no music) | 3.64 | Not significant | 0.97 | 6 | 2 |
| Story (music) | 4.07 | | 0.70 | 6 | 3 |

In DAS (see Table 2), the average password length with music was 10.43 and without, 13.79. The standard deviation of password length with music was 6.41, compared to 6.39 without. A t-test yields a result of t=1.34, P<0.20(two tails), indicating that the password length in the music group was significantly shorter than that in the control group. The background music increased the stroke count of passwords on average, but not to a statistically significant level. The standard deviation with respect to stroke count was higher with music (2.25 vs. 1.71).

While background music reduced the password length in DAS, it increased the password lengths in PassPoints and Story. As shown in Table 3, the average password length with music in PassPoints was 4.5 as opposed to 3.79 without. A t-test yields a result of t=1.61, P<0.20(two tails), indicating that there was statistically significant difference between two conditions. In Story, the password length for two groups differed by 0.43 (4.07 vs. 3.64), which is not statistically significant.

As such, the background music had a negative effect on DAS password length, but encouraged people to choose more complex passwords in PassPoints and Story.

# 5 Discussion

## 5.1 Validation of Hypotheses

Based on the results of our study, we now revisit our hypotheses that background music could improve humans' memory and then induced people to choose more complex passwords. This hypothesis was only supported in PassPoints. In PassPoints, people in the music condition not only chose significantly more complicated passwords, but also had significantly higher recall success rates in the long-term test. However, in DAS, the average password length of the music group was much shorter than that of the control group.

## 5.2 Recall Errors

This subsection will discuss the recall errors in DAS and PassPoints (Few errors occurred in Story and thus be ignored). People committed different types of error shown in Figure 4 (DAS) and Figure 5 (PassPoints). In DAS, there are three types of error: Stroke (i.e., entering more or less strokes), Pwd-Len (i.e., people could recall stroke count but forget the length of password) and Position (others including mixing up the stroke order or crossing incorrect cells). From figure 4, we can see that errors in Stroke and Pwd-Len account for the main proportion of recall errors. At the same time, music group committed more errors than the non-music group and the difference resulted possibly from the long-term recall test.

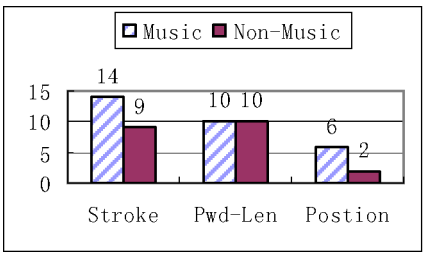

**Fig. 4.** Recall errors in DAS

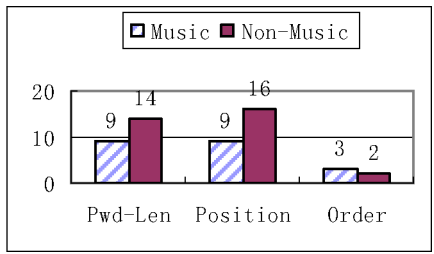

**Fig. 5.** Recall errors in PassPoints

There are also three types of error in PassPoints: Pwd-Len (i.e., forgetting the password length), Position (i.e., people can recall the password length but click points outside the tolerance region) and Order (i.e., only mixing up the click-points order) (see Figure 5). In this scheme, the nature of many recall failure was down to either forgetting the password length or clicking points outside the tolerance region. In recall errors and especially in Position errors, music group had a great advantage over non-music group, probably due to its higher success recall rate in the long-term recall test.

### 5.3 Limitations

Our intent in this study was to examine the effects of background music on the memorability of graphical passwords. We made our study follow the established methods of experimental psychology as much as possible and acknowledged that it did not mirror real-life usage. First, the participants in our study (all of them were university students of computer science) only represented a small part of the whole. It was important to get a good selection of people with various backgrounds in the further studies. Second, users are unlikely to familiar with and create three different graphical passwords one after the other in real life, or be asked to recall in quick succession them after one week (without having used any of them in the intervening time). Third, the participants had no incentive to perform as if protecting or accessing anything of real-life value to them, therefore it was not difficult to understand that many passwords created in both conditions were weak. For example, in Story, the average password length of the control group was less than 4. Furthermore, the effect of the background music volume remains to be discussed when it was embedded into a scheme. Despite these limitations, our controlled laboratory experiment paved the road to numerous further studies.

## 6 Conclusion

Results of the user study have shown that it is an effective enhancement to introduce baroque music to the PassPoints scheme. Surrounding with music, people not only tended to construct significantly more complicated passwords than their counterparts without the music stimulus, but also performed significantly better in terms of recall success in the long-term tests. This result indicated that the background music improved the memorability of passwords in PassPoints.

In DAS and Story, the introduction of background music has been shown unnecessary for security and usability. The recall of the passwords in both conditions was not statistically different from each other in short-term or long-term test. Furthe more, the background music significantly impaired the complexity of DAS passwords.

Although results obtained in three representative schemes are not consistent and should be treated with caution, we believe that this work provides a significant extension to the study of security and usability of graphical passwords. The future work includes a larger sale of studies with careful experimental design and Locimetrics systems will be our focus.

**Acknowledgments.** The authors would like to thank the reviewers for their careful reading of this paper and for their helpful and constructive comments. Project 60903198 supported by National Natural Science Foundation of China.